\documentclass[aps,pre,twocolumn,superscriptaddress,floatfix]{revtex4}
\usepackage{graphicx}
\usepackage{amsmath,amssymb}
\usepackage[utf8]{inputenc}
\usepackage{psfrag}
\usepackage{float}
\newcommand{\tkt}{T_{\scriptstyle\rm KT}}
\newcommand{\tc}{T_{\scriptstyle\rm c}}

\newcommand{\dmult}{\Delta_{\scriptstyle\rm mult}}

\newcommand{\rhov}{\rho_{\scriptstyle\rm v}}

\begin{document}
\title{Competing nematic interactions in a generalized XY model in two and three dimensions}

\author{Gabriel A. Canova}
\affiliation{Instituto de F\'\i sica, Universidade Federal do 
Rio Grande do Sul, CP 15051, 91501-970 Porto Alegre RS, Brazil} 

\author{Yan Levin}
\affiliation{Instituto de F\'\i sica, Universidade Federal do 
Rio Grande do Sul, CP 15051, 91501-970 Porto Alegre RS, Brazil}

\author{Jeferson J. Arenzon}
\affiliation{Instituto de F\'\i sica, Universidade Federal do 
Rio Grande do Sul, CP 15051, 91501-970 Porto Alegre RS, Brazil}

\date{\today}

\begin{abstract}
We study a generalization of the XY model with an additional nematic-like 
term through extensive numerical simulations and finite-size techniques,
both in two and three dimensions. While the original model favors local
alignment, the extra term induces angles of $2\pi/q$ between neighboring
spins. We focus here on the $q=8$ case (while presenting
new results for other values of $q$ as well) whose phase diagram is much
richer than the well known $q=2$ case. In particular, the model presents
not only continuous, standard transitions between Berezinskii-Kosterlitz-Thouless
(BKT) phases as in $q=2$, but also infinite order transitions involving
intermediate, competition driven phases absent for $q=2$ and 3. 
Besides presenting multiple transitions, our results show that
having vortices decoupling at a transition is not a suficient condition
for it to be of BKT type. 
\end{abstract}

\maketitle

\section{Introduction}
\label{section.introduction}

Two dimensional models with $U(1)$ group symmetry and isotropic, short range
interactions do not
present a standard second order transition as a consequence of the
Mermin-Wagner theorem~\cite{MeWa66}. Indeed, fluctuations (Goldstone modes) destroy
any long range order, even at low temperatures. Nonetheless, 2d models such as
the XY do present two different phases, separated by an
infinite order phase transition at $\tkt$ known as the Berezinskii-Kosterlitz-Thouless (BKT)
transition~\cite{Berezinskii71,KoTh74,KoTh81}. The low temperature phase is characterized by
bound pairs of vortices and antivortices and power-law decaying correlations
driven by the spin waves, 
while above the critical temperature the vortices decouple and
correlations decay exponentially. Moreover, unlike the usual thermodynamic 
phases, the low-temperature, quasi-long-range order BKT phase is critical at
all temperatures below $\tkt$. At this temperature, the helicity
modulus, which is the order parameter that measures how the system
responds to a global twist~\cite{FiBaJa73,NeKo77,OhJa79}, has a universal,
discontinuous jump signaling the decoupling of vortices and anti-vortices.

Here we study a generalization of the XY model with a
competing term that favors a different alignment angle, depending
on the parameter $q$, ${\cal H}=\sum_{\langle i j \rangle} U(\theta_i-\theta_j)$
where
\begin{equation}
U(\phi) = - \Delta\cos\phi -  (1 - \Delta) \cos(q\phi)
\label{eq.H}
\end{equation}
with $0\leq\Delta\leq 1$.
The sum is over nearest neighbors, $0\leq\theta_i< 2\pi$ and the usual XY model, with ferromagnetic
interactions is recovered either when $\Delta=1$ or $q=1$. For $\Delta=0$ the pure nematic term
induces skewed alignments with angles $2k\pi/q$, where $k\leq q$ is an integer. Since the partition
function for the cases $\Delta=0$ and 1 can be mapped onto each other by the transformation 
$q \theta_i \rightarrow \bar \theta_i$, the critical temperature~\cite{Hasenbusch05} is the same, 
$\tkt(\Delta=0)=\tkt(\Delta=1)\simeq 0.893$.
This transition, from the high temperature paramagnetic 
phase to a phase where there is local, non long ranged ordering, either nematic 
(for small $\Delta$) or ferromagnetic (large $\Delta$), is a BKT transition.
In the intermediate region where both terms compete, although new phases may appear
at low temperature, the transition from the paramagnetic phase seems to be BKT for all values 
of $\Delta$ (albeit it remains possible that a non BKT transition may exist close to
the multicritic point~\cite{ShLaFe11,HuWe13}). Interestingly, since the presence of a competing term helps to disrupt both
the nematic and ferromagnetic orderings, the transition temperature $\tkt(\Delta)$
is smaller at intermediate values of $\Delta$. At its minimum, that gets closer
to $\Delta=0.5$ as $q$ increases, several transition lines meet at a multicritical
point, $\dmult$. The parameter $\Delta$ can tune the relative strength of the two 
terms in Eq.~(\ref{eq.H}) and, as a consequence, which type of vortices, integer
or semi-integer (see below) is going to be relatively suppressed.
This class of models, with ferromagnetic and/or antiferromagnetic interactions, was used to
model the interlayer interactions of stacked bent-core molecules in liquid 
crystals~\cite{MaSaRo16}, DNA packing~\cite{Grason08}, structural phases of cyanide
polymers~\cite{CaClPaDaTuCoGo16,Zukovic16}, quasicondensation in atom-molecule, 
bosonic mixtures~\cite{BoWe12,BhEjEsFeHoSi12,PaRaRo16} and out-of-equilibrium
self-propelled polar particles~\cite{NgGiCh12}, with a similar, albeit dynamical, phase 
diagram in the latter case.

For $q=2$, in 2d, there are two transitions for 
$\Delta<\dmult$~\cite{KrJa54,Korshunov85,Korshunov86,LeGr85,LeGrTo86,CaCh89,Teixeira98,ShLaFe11,DiHl11,HuWe13}: as $T$ decreases there is
first a BKT transition to a phase with a local nematic ordering and, at a lower temperature,
a symmetry breaking transition in the Ising universality class to a phase with a local 
ferromagnetic alignment. In spite of the presence of a standard, continuous
transition, both phases are critical at every temperature,
but differ by the nature of the coupled topological defects they contain (see below). On the other
hand, for $\Delta>\dmult$, there is a single BKT transition
from the paramagnetic phase. A similar phase diagram was obtained in 3d~\cite{LuCh12},
but while the transition from the nematic to the ferromagnetic phase is still Ising like (3d),
the transition to the paramagnetic phase, for all $\Delta$, is continuous 
with exponents belonging to the 3dXY universality class (under different conditions, 
a BKT transition in 3d may occur as well~\cite{JoAlHoMaOpRe77,HoSh77,AmElRaSa82,AlJaSa86}). 
A closely related class of
models consists on a double XY model with an extra term coupling the two variables that
in the limit of strong coupling recovers the Hamiltonian 
Eq.~(\ref{eq.H})~\cite{GrKo86,ShGh06,PaRaRo16}.
The models with $q>2$ have been recently investigated as well~\cite{PoArLe11,CaLeAr14} (higher
harmonics have been also considered in Ref.~\cite{Himbergen86} in a related albeit
different model). The overall
phase diagram is similar for both $q=2$ and 3, the main difference being 
that for $q=3$ the transition between the ferromagnetic-like ({\bf F}$_0$) and the nematic-like 
({\bf N}) phase is in the 3-states Potts universality class (for all $q$, the {\bf N} phase
is unstable at low enough temperature if $\Delta\neq 0$). Moreover, for all values of
$\Delta$, the temperature at which the BKT transition occurs obeys~\cite{CaLeAr14} 
the lower bounds obtained by Romano~\cite{Romano06}. A related coupled XY model has been
studied, based on Ref.~\cite{BrAe82}, for $q=3$ both in 2d~\cite{GrKo86,JiHuKoStJiHu93} and 3d~\cite{GhSh05}
and for $q=6$ on a triangular lattice~\cite{ShJi05}.

For $q=8$~\cite{PoArLe11}, a representative large value of $q$, several new features are present as 
the former pseudo-ferromagnetic phase seems to split into several
regions with different quasi-long range ferromagnetic orderings, {\bf F}$_0$, {\bf F}$_1$ 
and {\bf F}$_2$ (some hints of an extra phase appear in an earlier study~\cite{ShJi05} 
of coupled XY models with a on-site coupling inducing a $q=6$ order), see Fig.~\ref{diag2Dq8}. 
Simulations with small lattice sizes~\cite{PoArLe11} were consistent with
the {\bf F}$_1$-{\bf F}$_0$ and {\bf F}$_2$-{\bf F}$_0$ transitions being in the BKT universality 
class. Interestingly, for some values of $\Delta$, there are two BKT transitions as the 
temperature is lowered.
The results of Ref.~\cite{PoArLe11} for $q=8$ were preliminary and some were not conclusive.
Indeed, the lattice sizes used were too small to confirm 
the 2d Ising universality class of the transition {\bf F}$_1$-{\bf F}$_2$ and the evidence
for the existence of the transition {\bf F}$_1$-{\bf F}$_0$ was admittedly quite weak. Therefore, 
this work is aimed not only to solve these issues, providing further data supporting or
clarifying the previous existence and universality claims, but also to extend to 3d the results 
for both $q=3$ and 8. In addition, other interesting questions remain open. How, as $q$
increases, the phase diagram changes from two to four (locally) ordered phases? The nematic
term, albeit with continuous variables, has some similarity with the discrete
Clock model. How similar is the behavior of both models? In order to answer these
questions, we present new results for larger lattices (on the square and cubic
lattices with periodic boundary conditions) also exploiting the power
provided by GPU computation and cluster algorithms~\cite{SwWa87,Wolff89}.

In order to clarify the nature of the phase transitions, we consider the modulus of the
generalized magnetization, 
\begin{equation}
m_k = \frac{1}{N}\left| \sum_i \exp(ik\theta_i)\right| 
\label{eq.magnetization}
\end{equation}
and the corresponding susceptibilities and Binder cumulants~\cite{Loison99,Hasenbusch08}
\begin{align}
\chi_k &= \beta N (\langle m_k^2 \rangle-\langle m_k\rangle^2) \label{eq.susc}\\
U_k    &=  \frac{\langle m_k^2 \rangle^2}{\langle m_k^4 \rangle} \label{eq.binder},
\end{align}
where $1\leq k\leq q$, $N=L^d$ and $\langle\ldots\rangle$ means thermal average. 
The specific heat is also measured in order to obtain a rough location of the
transition lines on the phase diagram.
In 3d, because the low temperature phase has genuine long-range order, or for
second order transitions that are present in 2d, the critical exponents 
$\beta$, $\gamma$ and $\nu$ may be obtained via standard
finite size scaling relations, $m = L^{-\beta/\nu} f(t L^{1/\nu})$ and 
$\chi = L^{\gamma/\nu} g(t L^{1/\nu})$ with $t\equiv T/\tc-1$.
In 2d, however, because the pseudo-ordered phase is critical everywhere,
the magnetization goes to zero while the susceptibility diverges
in the thermodynamic limit for all temperatures below the BKT transition. 

For a BKT transition, the proper order parameter is the helicity 
modulus~\cite{FiBaJa73,NeKo77,MiKi03}, the response of the system upon a small 
overall twist $\tau$ of spins in a particular direction. It is defined as
$\langle\Upsilon\rangle \equiv \left.\partial^2F/\partial\tau^2\right|_{\tau=0}=  
\langle e\rangle -N\beta \langle s^2\rangle$, 
where $F$ is the free energy,
$e\equiv N^{-1} \sum_{\langle ij\rangle_x} U_{ij}''(\phi)$ and  $s\equiv N^{-1} \sum_{\langle ij\rangle_x} U_{ij}'(\phi)$
(the sum is over the nearest neighbors along the direction of the twist), $\phi=\theta_i-\theta_j$ and $U_{ij}(\phi)$
is the potential between spins $i$ and $j$. Following Ref.~\cite{HuWe13}, for the Hamiltonian 
Eq.~(\ref{eq.H}),
\begin{align}
\Upsilon &= \frac{1}{N} \sum_{\langle ij\rangle_x} \left[ \Delta\cos\phi+q^2(1-\Delta)\cos(q\phi)\right] \nonumber \\
         &- \frac{\beta}{N}  \left(\sum_{\langle ij\rangle_x}\left[\Delta\sin\phi+q(1-\Delta)\sin(q\phi)\right]\right)^2.
\label{eq.helicity}
\end{align}
To improve the accuracy, we average $\Upsilon$ both along the horizontal and vertical
directions. Moreover, a fourth-order helicity modulus $\Upsilon_4$ can be 
introduced in a similar way~\cite{MiKi03},  
$\langle\Upsilon_4\rangle \equiv \left.\partial^4F/\partial\tau^4\right|_{\tau=0}$,
with the perturbed free energy, up to fourth order, given by
$F(\tau)\simeq\langle\Upsilon\rangle\tau^2/2! + \langle\Upsilon_4\rangle\tau^4/4!$. 
The BKT theory predicts, for the original XY model, 
that the helicity modulus is zero within the disordered phase and  
jumps to a finite value at the transition to the ordered phase, where 
the critical temperature is given by the 
condition $\Upsilon(\tkt)=2\tkt/\pi$~\cite{FiBaJa73,NeKo77}. This transition is 
driven by the decoupling of pairs of integer vortices and anti-vortices. 
Since the critical temperature must be the same for both $\Delta=0$ and 1, the condition becomes 
$\Upsilon(\tkt)=2\tkt/\lambda^2\pi$, where $\lambda=1/q$ is the charge of the vortex.
The $q^2$ factor is introduced because Eq.~(\ref{eq.helicity}), for $\Delta=0$ and 1,
differs by this factor. For $q=2$, the topological excitations in the nematic phase (small 
$\Delta$) are the {\it half}-vortices, with related charge $\lambda=\pm 1/2$~\cite{LeGr85,CaCh89,Korshunov85,HuWe13}. 
For $q=3$, vortices excitations were recently found, related to a 
$\lambda=\pm 1/3$ charge~\cite{CaLeAr14,PoArLe11} in the nematic phase. For larger $q$, however,
since several new phases may be present (see below),
it remains unclear what kind of topological excitation each phase does have 
and which is the nature of each transition. It is possible to somewhat
characterize the vortices through the winding number, obtained by summing the
phase difference $\phi=\theta_i-\theta_j$ counter-clockwise around every site, 
including all nearest neighbors and taking
care that $|\phi|\leq\pi$~\cite{ToCh79}. 
For an integer vortex, this sum is $\pm 2\pi n$ and the
density of vortices is $\rho_{\scriptstyle\rm v}\equiv N_{\scriptstyle\rm v}/N$, 
where $N_{\scriptstyle\rm v}$ is the total number of vortices (that may also be
distinguished by their sign). This is also easily generalized to semi-integer
vortices.

The paper is organized as follows. In Sect.~\ref{results.2d} we present an
improved analysis of the $q=8$ case in 2d with larger lattices and additional
observables~\cite{PoArLe11} sometimes presenting,
for the sake of comparison, results for other values of $q$ as well. 
Then, Sect.~\ref{results.3d}
shows the results for both $q=3$ and 8 in 3d. In Section~\ref{conclusion} we 
discuss these results and present our conclusions.

\section{Results}
\label{results}

\subsection{2d}
\label{results.2d}

\begin{figure}[ht]
\includegraphics[width=8cm]{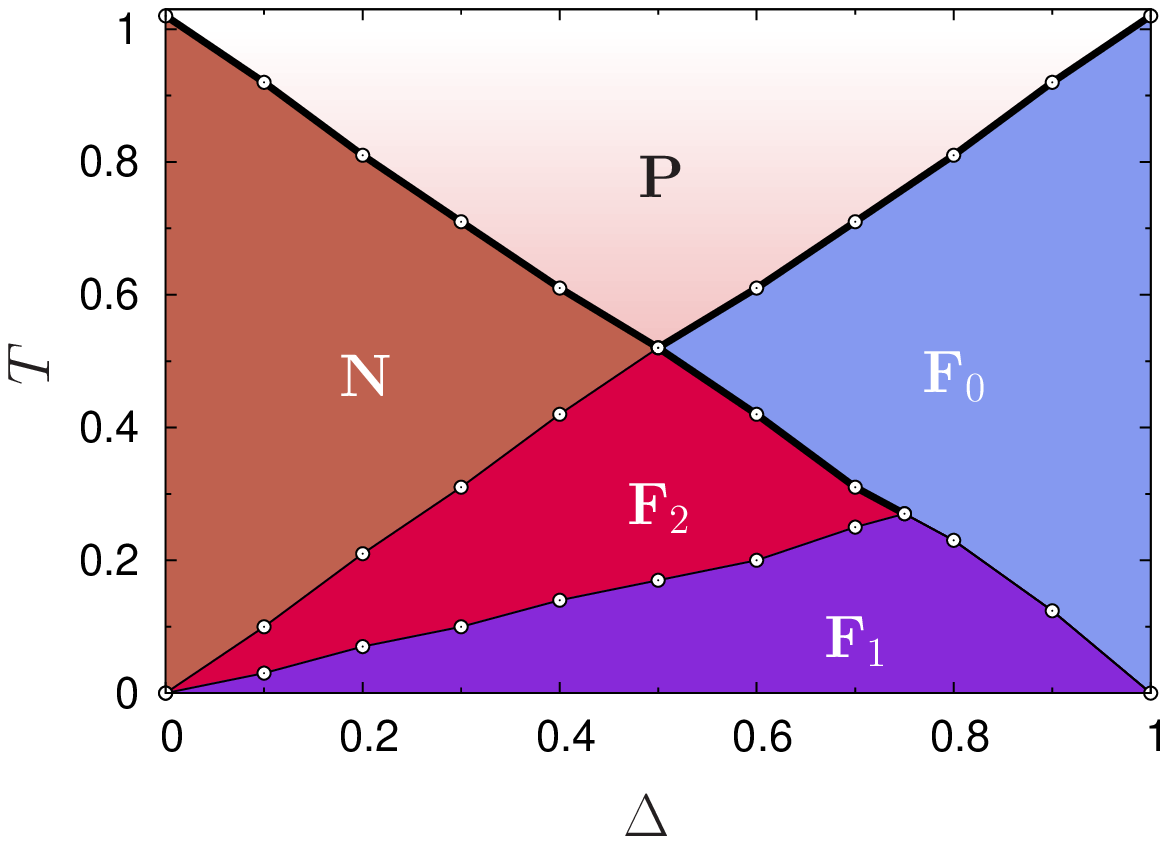}
\includegraphics[width=8cm]{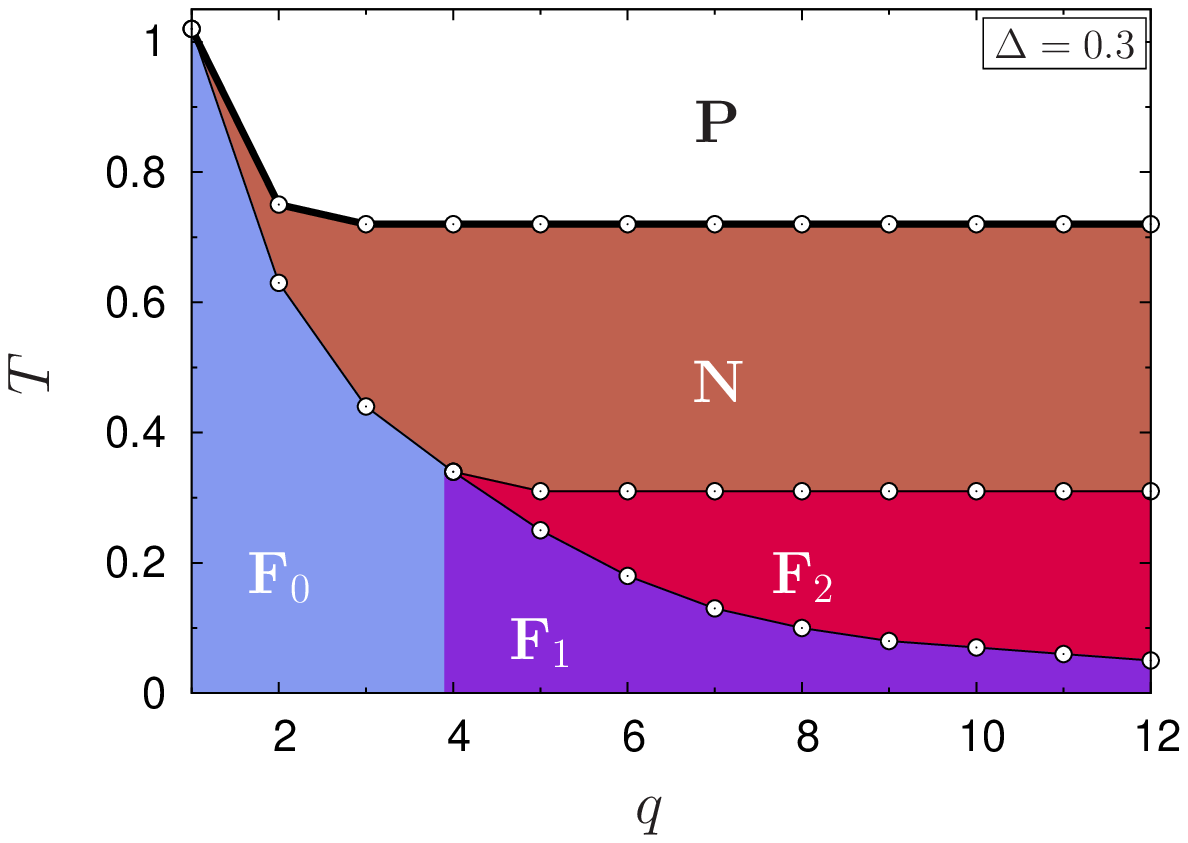}
\caption{(Top panel) Qualitative phase diagram for the 2d generalized
XY model with $q=8$. The points correspond to the maxima of the specific heat
for $L=64$ 
while the lines (thick/thin for BKT/second order transitions) 
are just a guide to the eyes. 
(Bottom panel) Transition lines for several values of $q$ for $\Delta=0.3$. There is a bifurcation 
of the {\bf N}-{\bf F}$_0$ transition for $q>4$, similar to what happens in the Clock 
model with $\mathbb{Z}_q$ symmetry. The lowest transition line 
decreases as $q^{-2}$.} 
\label{diag2Dq8}
\end{figure}

The phase diagram has the same topology for both 
$q=2$~\cite{Korshunov85,LeGr85,CaCh89,HuWe13} and 3~\cite{PoArLe11,CaLeAr14}. 
Besides the paramagnetic
phase ({\bf P}) at high temperatures, there are two phases with quasi long range order, 
each one associated with the pure cases at $\Delta=0$ and 1. The former 
is a nematic-like phase ({\bf N}) while the latter has local ferromagnetic
ordering ({\bf F}$_0$). The {\bf N}-{\bf F}$_0$ transition is second order
and is either in the Ising or in the
three states Potts model universality class for $q=2$ and 3, respectively. Ref.~\cite{PoArLe11} 
also considered the $q=8$ case in which the region previously occupied by
{\bf F}$_0$ separates in three phases, 
all having local
ordering similar to the ferromagnetic state (to be discussed in detail
below). 
Through the position of the specific heat peak, we obtain a rough estimate of the
phase boundaries for several values of $q$ (not being very precise, the
transition line is somewhat displaced). For $q=2$ and 
3 the  {\bf N}-{\bf P} BKT transition runs very close to the $\tkt(0)(1-\Delta)$ line,
the lower bound for the critical temperature predicted in Ref.~\cite{Romano06}, 
and ends at the multicritical point ($\dmult$) where several transition lines meet. 
In these two cases, on the other hand, since the multicritical point is still far 
from $\Delta=0.5$, the {\bf F}$_0$-{\bf P} BKT transition obbeys, but is not so close to the
corresponding lower bound, $\tkt(\Delta)\geq \tkt(0) \Delta$, for $\Delta\geq\dmult$. 
An important,
yet open, issue is how this complex structure unfolds as $q$ increases.
As $q$ increases, $\dmult$ approaches 0.5 and both BKT transitions to the {\bf P}
phase roughly follow those lower bounds (see, e.g., the thick border of phase
{\bf P} in Fig.~\ref{diag2Dq8}). A new transition line appears for $q=4$,
extending from the multicritical point down to the corner at $\Delta=1$ and $T=0$,
dividing the {\bf F}$_0$ phase in two, with a new phase, {\bf F}$_1$, being
created below both {\bf N} and {\bf F}$_0$ for all $0<\Delta<1$. 
For $q=5$, the {\bf N}-{\bf F}$_1$ transition splits
in two, creating another intermediate phase, {\bf F}$_2$: there is local alignment
along several directions, as in the {\bf N} phase, all of them belonging to the 
same half-plane, as in the {\bf F}$_1$ phase. Along with
that, the multicritical point also bifurcates, originating a new point where
all the {\bf F}$_i$ phases meet. 
Differently from the {\bf N} and {\bf F}$_0$ phases that are driven,
respectively, by the pure first and second terms in Eq.~(\ref{eq.H}), 
the new phases, {\bf F}$_1$ and {\bf F}$_2$, are driven by the competition 
between these two terms and, as will be seen below, have a mixture
of the topological defects that characterize both {\bf N} and {\bf F}$_0$.
The region occupied by the new phase, {\bf F}$_2$, increases with $q$
and the critical temperature at the  {\bf F}$_1$-{\bf F}$_2$ border
decreases as $q^{-2}$. These scenarios are summarized in the
bottom panel of Fig.~\ref{diag2Dq8} for $\Delta<\dmult$ and we remark
the resemblance with Fig.~1 from Ref.~\cite{LaPfWe06} for the Clock model, 
whose symmetry is discrete. 
Indeed, a similar sequence of phase splitting transitions
occurs in that model~\cite{JoKaKiNe77,KuHuToOs13}. 
The two transitions, from the para to the ferromagnetic phase in the Clock
model and {\bf N}-{\bf F}$_2$ here, are in the Ising universality class for $q=2$ and 4, but belong to the three states Potts model for $q=3$.
Moreover, while here a new phase ({\bf F}$_2$) appears for $q>4$,  in
the Clock model, a similar, intermediate phase with coarse grained $U(1)$ symmetry 
(and BKT nature) appears as well, between the paramagnetic and the low temperature, 
ferromagnetic phase. In both models, the transition temperature to the lowest
temperature phase decreases as $q^{-2}$ and, as $q\to\infty$,
this phase shrinks and disappears. By suppressing {\bf F}$_1$ in this limit,
only four phases remain in the phase diagram once again. 
For $q>5$, remarkably, while in the Clock model, whose spins are discrete,
the two transitions are BKT, here, at least for those values of $q$ that we studied, 
the transitions from {\bf F}$_2$ to both {\bf N} and {\bf F}$_1$ seems to be of second order.
We now describe in detail the nature of the phases and transitions for 
$q=8$.

\begin{figure}[ht]
\includegraphics[width=8cm]{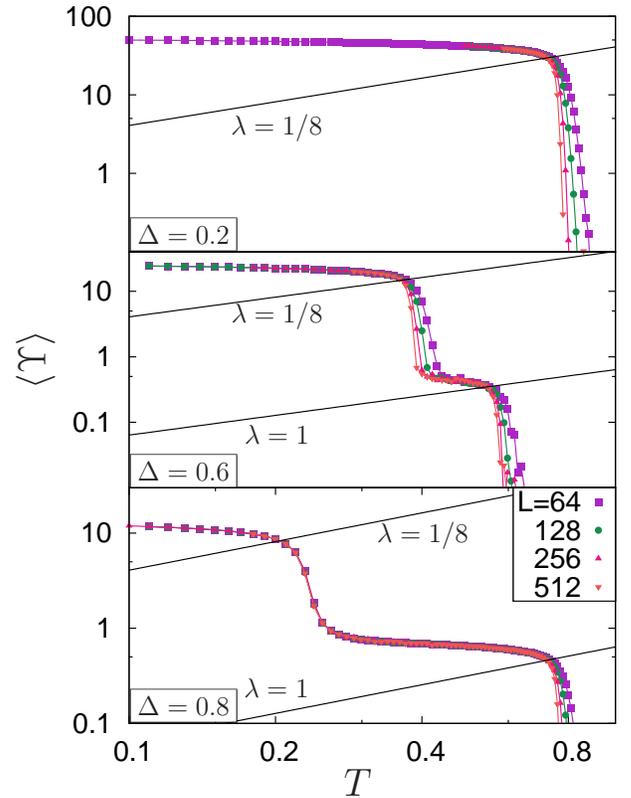} 
\caption{Helicity modulus {\it vs} temperature, in log-log scale, for the 2d, 
$q=8$ model with $\Delta=0.2$ (top panel), 0.6 (middle panel) and 0.8 (bottom panel).
The crossing points of 
$\langle\Upsilon\rangle$ with the straight lines $2T/\lambda^2\pi$ 
with $\lambda=1$ or 1/8 mark the putative transitions. For $\Delta=0.2$, 
at the {\bf N}-{\bf P} transition, $\lambda=1/8$. For $\Delta=0.6$
there are two BKT transitions with increasing temperatures, {\bf F}$_2$-{\bf F}$_0$  and
{\bf F}$_0$-{\bf P}, with $\lambda=1/8$ and 1, respectively. Both jumps 
are size dependent while $\langle\Upsilon\rangle$ has no jump at the transition 
{\bf F}$_1$-{\bf F}$_2$
occurring at a lower temperature. When $\Delta=0.8$, only for the {\bf F}$_0$-{\bf P} 
transition there is an observable size dependence of $\langle\Upsilon\rangle$. 
At $T\simeq 0.2$, there is a large, but smooth change without any perceptible 
size dependence (within the range considered here). 
In all cases, the helicity vanishes at the transition to the high 
temperature phase.}
\label{fig.helicity} 
\end{figure}

\begin{figure}[ht]
\includegraphics[width=8cm]{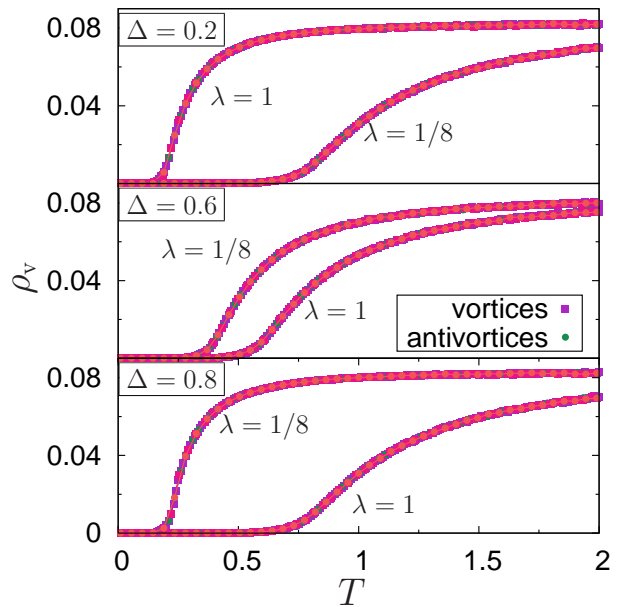} 
\caption{Density of vortices and antivortices  {\it vs} temperature for 
the 2d, $q=8$ model for several values of $\Delta$. The behavior is size 
($L=64$ and 256) and vorticity (positive or negative) independent
since all sets collapse onto each other. As $T\to\infty$, all
curves approach the asymptotic value 1/12~\cite{JeWe92} as $T^{-2}$.}
\label{fig.vortex_density} 
\end{figure}
\begin{figure}[ht]
\includegraphics[width=8cm]{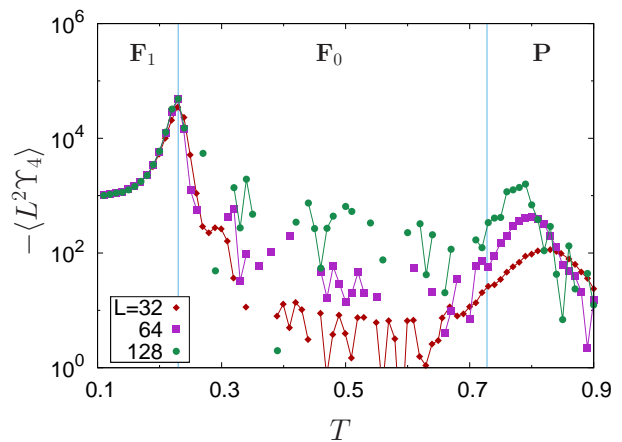}
\caption{High order helicity for the 2d, $q=8$ case with $\Delta=0.8$ at the
{\bf F}$_1$-{\bf F}$_0$ (left) and {\bf F}$_0$-{\bf P} (right) transitions. 
In the intermediate {\bf F}$_0$ phase the results are noisier while both at 
the paramagnetic and {\bf F}$_1$ phases these
fluctuations are much suppressed either by the smaller correlation between spins
or the larger stiffness of the system, respectively.}
\label{hel4D08}
\end{figure}

Fig.~\ref{fig.helicity} shows the behavior of the helicity modulus 
$\langle\Upsilon\rangle$ for three vertical cuts 
of the phase
diagram. In the thermodynamical limit, a BKT transition is signaled by a discontinuous
jump of the helicity from $\Upsilon(\tkt)=2\tkt/\lambda^2\pi$ to 0. For a finite system, it is expected~\cite{WeMi88} that the critical temperature 
reaches logarithmically its asymptotic value as the system size increases, 
\begin{equation}
\Upsilon_{\scriptstyle\rm fit}(L)=\frac{2TA}{\pi}\left( 1 + \frac{1}{2} \frac{1}{\log CL}\right),
\label{eq.logfit}
\end{equation}
where $A$ and $C$ are fitting parameters. Notice that in order to present such a
behavior, the curves must be size dependent close to the transition.
The parameter $A$, the vorticity, provides an estimate to $1/\lambda$ at the transition~\cite{HuWe13}.  Applying Eq.~(\ref{eq.logfit}), when appropriate, and minimizing the quadratic error as defined in 
Refs.~\cite{HuWe13,CaLeAr14}, we find either $A\simeq 64$ or 1, 
depending on the transition, corresponding to a $\lambda\simeq 1/8$
(fractional) or 1 (integer) charge, respectively. Fig.~\ref{fig.vortex_density} shows the  
density $\rhov$ of vortices and antivortices
for the same three values of $\Delta$ of Fig.~\ref{fig.helicity}. There seems to be
no size dependence as the curves for both $L=64$ and 256 perfectly match.
No distinction exists also between the number of vortices and antivortices: be either
free ou bound, their densities are always the same. 
Vortices unbind close to all transitions
shown in the phase diagram, as shown by an increase in 
the density of vortices, except for the {\bf F}$_1$-{\bf F}$_2$ one. 
Thus, both species of vortices, integer and fractional, remain bounded inside
{\bf F}$_1$ and {\bf F}$_2$ and when transitioning to either the {\bf N} or {\bf F}$_0$
phases one of the species unbinds, while only at the border with the {\bf P}
phase occurs the unbinding of the remaining vortices (where the helicity becomes
zero). In all cases, $\rhov$ monotonously increases, attaining a limiting
value when $T\to\infty$~\cite{JeWe92}. However, in this limit, because of the
strong thermal noise, no vortex exists and the finite result is an artifact
of the lattice discretization that uses a small number of sites around each
site in the definition of $\rhov$. This is similar to the geometric clusters,
group of nearest neighbors parallel spins, whose size distribution at $T\to\infty$ 
is an exponencial, but the physical, Coniglio-Klein clusters of
correlated spins correspond to single sites. Indeed, with the above definition of
$\rhov$, the probability of obtaining a (unitary winding number) vortex
purely by chance is (the same result applies for antivortices as well)
$$
P_4 = \frac{1}{(2\pi)^3}\int_0^{\pi}d\theta_2\int_{\theta_2}^{\theta_2+\pi} d\theta_3
\int_{{\text max}(\pi,\theta_3)}^{{\text min}(2\pi,\theta_3+\pi)}d\theta_4
= \frac{1}{12}
$$
where the angle $\theta_i$ is the state of each neighbour spin (and $\theta_1=0$).
This asymptotic value is approached as $|\rhov-1/12|\sim T^{-2}$ when
$T\to\infty$ (although the coefficients differ for integer and semi-integer vortices).
In an attempt to get closer to the continuous case, we may generalize the definition
and consider loops with $n$ sites around each spin. While for small $n$ this can be
done analitically, as $n$ increases one has to resort to numerical
evaluations. Combining these results, we conjecture that $P_n$ is given by
$$
P_n = \frac{1}{(n-1)!} \left( 1 - \frac{n}{2^{n-1}}\right).
$$
This expression decreases very fast as $n$ increases, since it become exponentially
more difficult to have a vortex by chance alone. This refined classification gives indeed a peaked 
density of vortices close to the transition but does not change the point at which
the vortices unbind. 
For all $\Delta<\dmult$ (top panel on Figs.~\ref{fig.helicity} and \ref{fig.vortex_density})
there is an {\bf N}-{\bf P} transition with a discontinuous jump of
the helicity (notice the size dependence that indicates a BKT transition) 
accompanied by the unbinding of the fractional charges, Fig.~\ref{fig.vortex_density}. 
Once in the paramagnetic
phase, the full $U(1)$ symmetry is recovered, vortices and antivortices
are no longer bound together and the helicity vanishes. 
Although not visible on the scale of Fig.~\ref{fig.helicity} (top panel), 
there is a tiny, size independent decrease of the helicity when the border 
{\bf F}$_2$-{\bf N} is crossed as well, indicating a non BKT transition. 
Nonetheless, the integer vortices decouple at this transition 
(Fig.~\ref{fig.vortex_density}). Inside the phases {\bf F}$_1$ and {\bf F}$_2$, 
both kinds of vortices are present and remain bound in vortex-antivortex pairs.
For $\Delta>\dmult$, the transition {\bf F}$_0$-{\bf P} involves the dissociation
of the integer charges, $\lambda\simeq 1$ (lower straight line in middle and 
bottom panels of Figs.~\ref{fig.helicity} and \ref{fig.vortex_density}). 
In this case, the fractional charges
decouple at a lower temperature as can be seen in Fig.~\ref{fig.vortex_density}.
The $\Delta=0.6$ case (middle panel) is an example with two consecutive discontinuous
decreases of the helicity, both associated with unbinding of vortices and BKT
transitions~\cite{PoArLe11} (notice the size dependence). At the first, lower 
$T$ transition ({\bf F}$_2$-{\bf F}$_0$), only the fractional vortices decouple 
and the helicity decreases to a smaller value corresponding to the integer vortices.
Notice that because of the $q^2$ factor in the definition of $\langle\Upsilon\rangle$, 
the contribution from fractional vortices is significantly higher than the one from 
the integer vortices. There is a similar decrease
in the bottom panel, without the size dependence characteristic of the BKT
transition, which hints to a non-BKT nature of the transition {\bf F}$_1$-{\bf F}$_0$. 
While the {\bf N} and {\bf F}$_0$ phases have either fractional
or integer bound vortices, respectively, the phases {\bf F}$_1$ and {\bf F}$_2$,
driven by the competition between the two terms in the potential, have mixed
charges, with both species of vortices coexisting and bound in vortex-antivortex
pairs. 
It is indeed because of this coexistence inside both phases {\bf F}$_1$ 
and {\bf F}$_2$, with no unbinding whatsoever at the transition,  
that the helicity does not show any particular feature as the border 
{\bf F}$_1$-{\bf F}$_2$ is crossed.
Summarizing the evidence gathered from the helicity and the density of vortices, 
besides the two transitions to the paramagnetic
phase, also the {\bf F}$_2$-{\bf F}$_0$ transition is BKT (all shown as thick
lines in the phase diagrams of Fig.~\ref{diag2Dq8}). 
Further evidence (not shown) is provided by the susceptibility, Eq.~(\ref{eq.susc}),
whose behavior is consistent with the one expected at a BKT transition: 
despite the absence of
long range order, finite systems still have a finite magnetization and 
a divergent susceptibility that scales, for finite systems, as
$\chi(\tkt)\sim L^{2-\eta}$, with $\eta= 1/4$ at the BKT 
transition and non universal values inside the low temperature phase. 
Similar information is conveyed in Fig.~\ref{hel4D08}, for $\Delta=0.8$, by the 
fourth order helicity modulus, $\langle L^2\Upsilon_4\rangle$,  that is
expected to diverge at a BKT transition~\cite{MiKi03}. We can see that the fourth order helicity 
increases only at the transition {\bf F}$_0$-{\bf P} while at the {\bf F}$_1$-{\bf F}$_0$
there is no apparent size dependence, again signalling the different nature of
both transitions.


\begin{figure}[ht]
\begin{center}
\includegraphics[width=8cm]{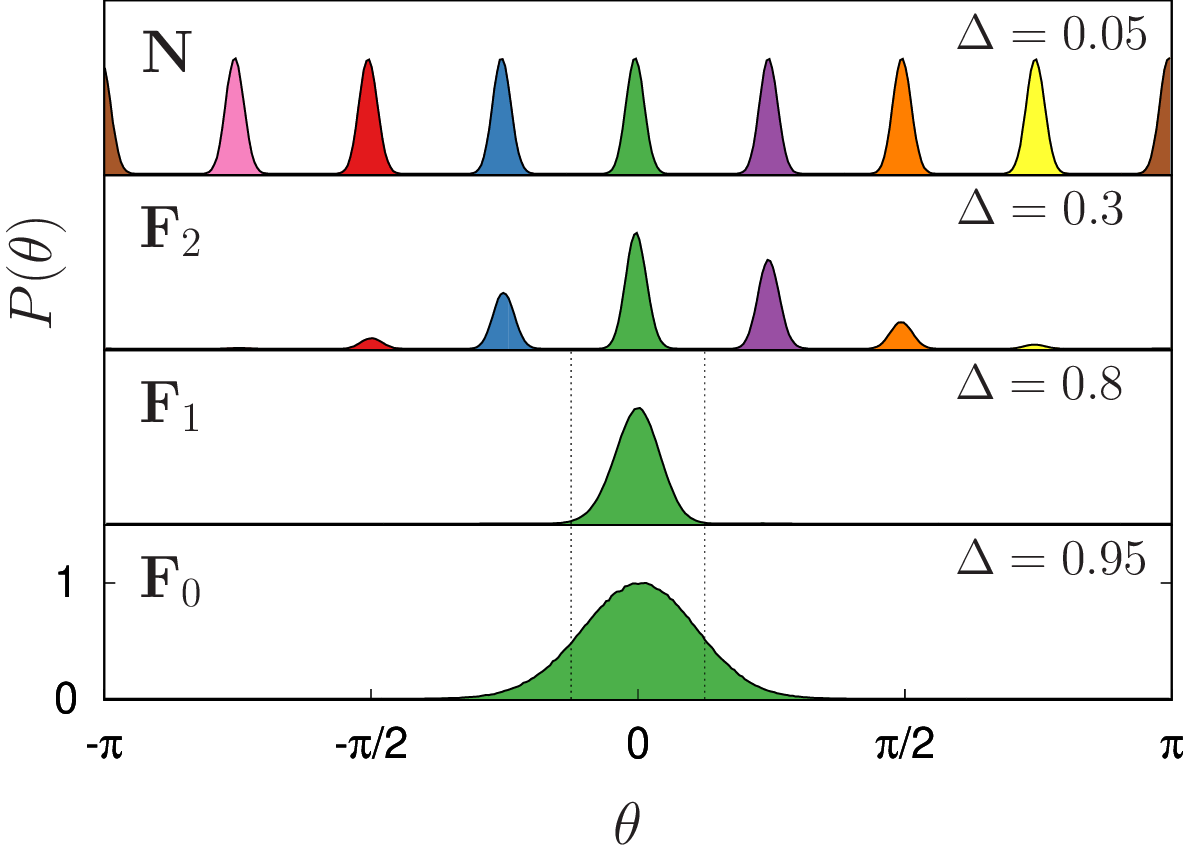}

(a)

\includegraphics[width=4cm]{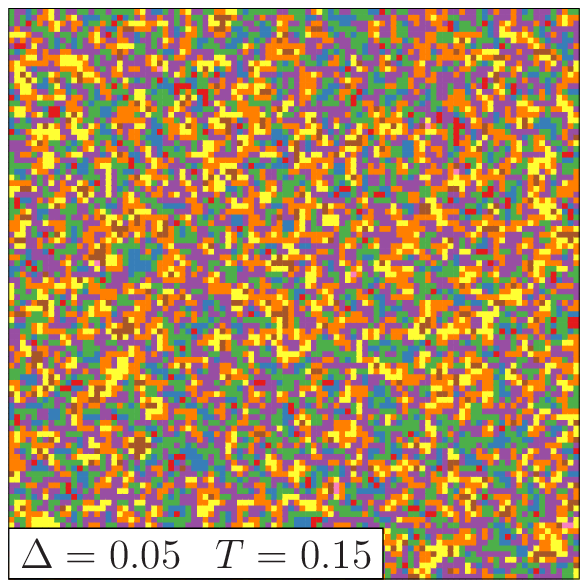}
\includegraphics[width=4cm]{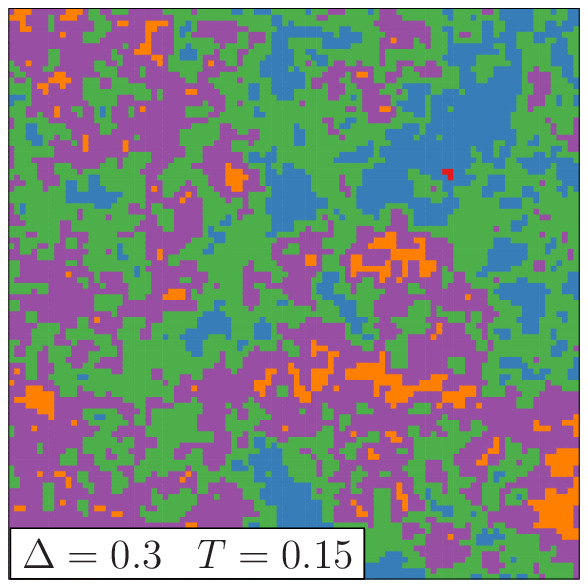}

(b)\hspace{3.5cm}(c)
\end{center}
\caption{(a) Distribution of orientations in a single configuration with
$L=1024$ and several values of $\Delta$ taken from a horizontal 
cut of the phase diagram at $T=0.15$, corresponding to the several
low temperature phases. The vertical dotted lines show that at the
{\bf F}$_1$-{\bf F}$_0$ transition, the width of the distribution
becomes larger than 1/8 of the circle. Snapshots of a $100^2$ region of typical
configurations in the {\bf N} (b) and {\bf F}$_2$ (c) phases. The color indicates
to which peak of the corresponding panel in (a) the spin belongs to.} 
\label{fig.histq8}
\end{figure}

\begin{figure}[ht]
\includegraphics[width=8cm]{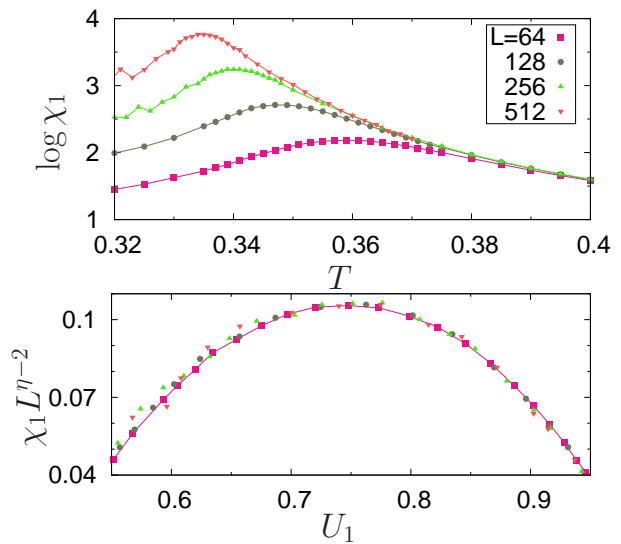}
\caption{The transition {\bf F}$_2$-{\bf N} in the 2d case for $q=8$ and $\Delta=0.35$. 
The susceptibility $\chi_1$ (top panel) as a function of temperature close
to the transition at $\tc\simeq 0.325$ (vertical line). 
(Bottom panel) Rescaled susceptibility {\it vs} the Binder cumulant for several system sizes. 
The data collapse is obtained with the Ising 2d exponent $\eta=1/4$.}
\label{X1D035}
\end{figure}

Having shown evidences that the transitions {\bf F}$_2$-{\bf N}, {\bf F}$_1$-{\bf F}$_2$
and {\bf F}$_1$-{\bf F}$_0$ are not BKT, we now describe the properties of these
lines, in particular to which universality classes  they belong to.
Fig.~\ref{fig.histq8}a shows, for a single configuration, the
distribution of the spins~\cite{PoArLe11}.
As shown in Ref.~\cite{PoArLe11}, the transition between {\bf F}$_2$ and {\bf N} 
corresponds to a reflection symmetry breaking transition in the Ising universality 
class where, from the eight preferential directions symmetrically disposed around 
the circle in the {\bf N} phase, only four remain, all in the same half-plane, after
the transition. This is shown in the top two panels while typical configurations 
for these two phases, in which each peak of the distribution was colored differently 
are shown in Figs.~\ref{fig.histq8}b and c. 
We notice in Fig.~\ref{fig.histq8}b that the ferromagnetic interaction, being small
for $\Delta=0.05$, does not build clusters with spins belonging to the same peak,
instead, neighboring spins tend to obbey the nematic term. For a larger $\Delta$, 
Fig.~\ref{fig.histq8}c, even if the temperature is slightly higher, the ferromagnetic
term increases the size of the clusters (notice also that the system is magnetized
in this case).
Since in phase {\bf F}$_2$ spins have a preferred direction, finite systems are magnetized, $m_1>0$, and 
its associated susceptibility, $\chi_1$, presents a divergence as the temperature
decreases towards the transition. Fig.~\ref{X1D035}, top panel, shows $\chi_1$ as a
function of the temperature for several system sizes for $\Delta=0.35$. At the
critical temperature, $\chi_1 \sim L^{1.753}$, consistent with the 2d Ising value, 
$2-\eta=7/4$~\cite{PoArLe11}. Inside the {\bf N} phase, because of the
circular symmetry, $\chi_1$ gives results equivalent to the
paramagnetic phase, while once in the phase {\bf F}$_2$ the critical nature
of this phase presents an always diverging susceptibility, but with a temperature
dependent, non universal exponent, and the curves do not collapse away from the
transition. The bottom panel shows a good data collapse of $\chi_1$ against the Binder 
cumulant~\cite{Loison99,Hasenbusch08}, Eq.~(\ref{eq.binder}), with no required
knowledge of the critical temperature. The scaling of the Binder cumulant, close to the transition, 
is $U_1=h(L/\xi)$, where $h(x)$ is a scaling function and $\xi$ is the correlation length. 
Also, since $\chi_1=L^{2-\eta}g(L/\xi)$, we expect that 
$\chi_1L^{\eta-2}=g(h^{-1}(U_1))$. When plotted as a function of the temperature
(not shown), the Binder cumulant indeed assumes (roughly) the same value for
different system sizes at the transition.  

\begin{figure}[ht]
\includegraphics[width=8cm]{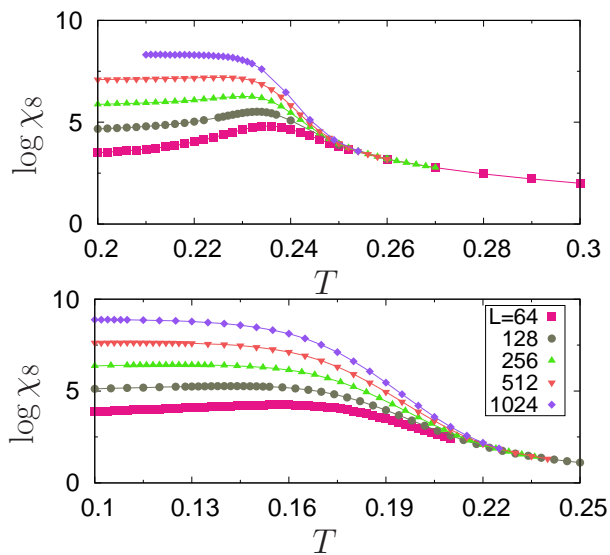}
\caption{Susceptibility {\it vs} temperature for the 2dXY model with $q=8$ at the {\bf F}$_1$-{\bf F}$_0$ transition for $\Delta=0.8$ (top) and $0.9$ (bottom). At the transition points, the susceptibility scales as: $\chi_8^*(\Delta=0.8)\sim L^{1.437}$ and $\chi_8^*(\Delta=0.9)\sim L^{1.741}$.}
\label{X8D08}
\end{figure}

Ref.~\cite{PoArLe11} presented preliminary data that seemed to indicate 
that the transition {\bf F}$_1$-{\bf F}$_0$ was also of BKT type. However, the 
behavior of the helicity discussed above hints at a second order transition. 
Indeed, in the region where one would expect a transition (close to the line 
$\tkt(0)(1-\Delta)$, e.g., $T\simeq 0.2$ for $\Delta=0.8$),  
there is a sudden change of the helicity, but it does not scale with the system size, indicating 
that in the thermodynamic limit the discontinuity associated with the BKT 
transition is not present. Nonetheless, fractional vortices unbind at this
transition, as seen by the behavior of $\rhov$ in the bottom panel of
Fig.~\ref{fig.vortex_density}.
In both {\bf F}$_1$ and {\bf F}$_0$ phases, 
see Fig.~\ref{fig.histq8}, the average distribution of orientations has a single peak that continuously becomes more narrow as the temperature decreases. Inside the phase 
{\bf F}$_0$, the variance of the distribution is larger than $\pi/8$, $m_8$ maps it 
onto the whole circle and does not distinguish it from the paramagnetic phase
(see bottom panels of Fig.~\ref{fig.histq8}a). 
The transition to the {\bf F}$_1$ phase occurs when the width becomes
smaller than this value and corresponds to a strong increase of $\chi_8$ at
the transition, as shown in Fig.~\ref{X8D08} for $\Delta=0.8$ and 0.9.
Notice that although the exponent of the susceptibility differs in both cases,
for $\Delta=0.9$ it is compatible with the Ising value.
Whether this discrepancy is real and the transition line has
non-universal, changing exponents, or whether it is caused by the proximity
to the second multicritical point where all three phases {\bf F}$_i$ meet, 
is still an open question and requires further simulations. 
Thus, the overall evidence points to a second order transition
between phases {\bf F}$_0$ and {\bf F}$_1$ compatible with the Ising 
universality class, at least on the rightmost part of such line.

\begin{figure}[ht]
\includegraphics[width=7.5cm]{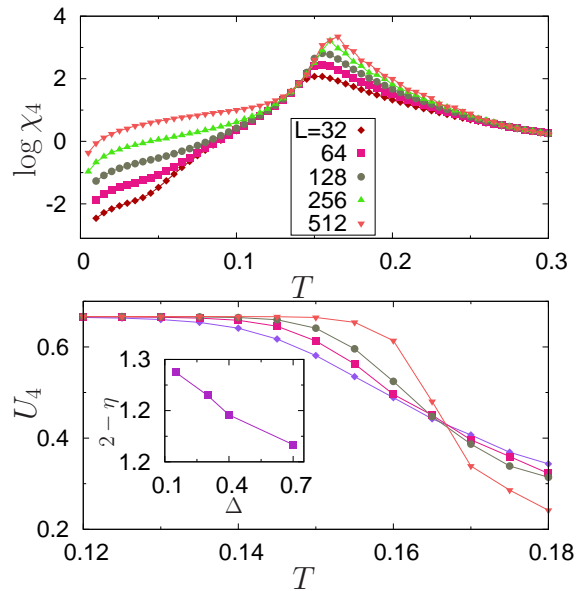}
\caption{(Top) Susceptibility $\chi_4$ {\it vs} temperature for $q=8$ at the {\bf F}$_1$-{\bf F}$_2$ transition for $\Delta=0.4$. Inset: the exponent of $\chi_4$ along the transition line showing a dependence, for the sizes considered here, on $\Delta$. (Bottom) Binder cumulant, for different
sizes, crossing in the region of the putative transition.}
\label{fig.f1f2}
\end{figure}

Finally, the transition between the phases {\bf F}$_1$ and {\bf F}$_2$ is the more
elusive one.
Differently from the other two similar transitions, 
in which one specie of vortices decouples despite being second order, 
here both species remain bound in vortex-antivortex pairs across the transition 
line, Fig.~\ref{fig.vortex_density}. The helicity is not able to detect the 
transition as there is no 
difference in the cost to produce an overall twist in the system since the
density of (anti)vortices, that are responsible for the stiffness, barely changes. 
Therefore, the best evidence we have for this transition comes from 
the susceptibility and the Binder cumulant. 
Fig.~\ref{fig.histq8} shows that the distribution of spin orientations is either
a single peak (phase {\bf F}$_1$) or several peaks of different heights
concentrated on a half-plane (phase {\bf F}$_2$). The appropriate order parameter in this
transition is thus $m_4$ and Fig.~\ref{fig.f1f2} shows the
corresponding fluctuations, $\chi_4$.  Notice that the
shape of these curves differs from the previous cases in which the plateau
corresponding to the critical nature of a BKT phase was much higher. Here, although such a
plateau seems to be developing, it is yet far from merging with the peak, what may also
indicate the presence of strong finite size effects. For different system sizes, in the 
same region where the peak increases, we also observe, Fig.~\ref{fig.f1f2} (bottom), the 
crossing of the Binder cumulant. The inset of Fig.~\ref{fig.f1f2} (top) shows that 
the critical exponent $2-\eta$ of $\chi_4$, as measured with the available sizes, 
seems to depend on $\Delta$,
continuously decreasing from, roughly, 1.28 to 1.21 along the {\bf F}$_1$-{\bf F}$_2$ line.
We do not rule out that larger system sizes and logarithmic corrections, 
when taken into account, may play a role either restoring universality or pointing
to a crossover instead of a transition. Otherwise, there may exist
an yet unexplored~\cite{AjVa09} connection with the Ashkin-Teller~\cite{AsTe43,GrWi81,BaReRi85,Goldschmidt86} or the eight vertex
model~\cite{Sutherland70,Baxter71,KaWa71}.

\subsection{3d}
\label{results.3d}

\begin{figure}[ht!]
\includegraphics[width=8cm]{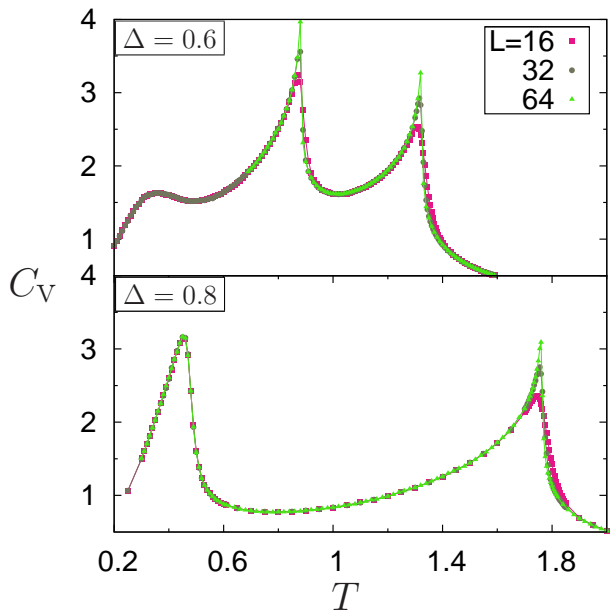}
\caption{Specific heat of the generalized 3dXY model for $q=8$ and $\Delta=0.6$ (top)
and 0.8 (bottom). Although several peaks are visible, those occurring at the lowest
temperature seem not to increase with the system size (see discussion in the text)
while the others have the characteristic form of a lambda transition, with a cusp at a finite
value instead of a divergence (negative $\alpha$).
The overall evidence indicates that these low $T$ peaks do not correspond to a phase transition, 
but to a crossover.}
\label{fig.3Dq8CvD06}
\end{figure}

\begin{figure*}[htb!]
\includegraphics[width=8cm]{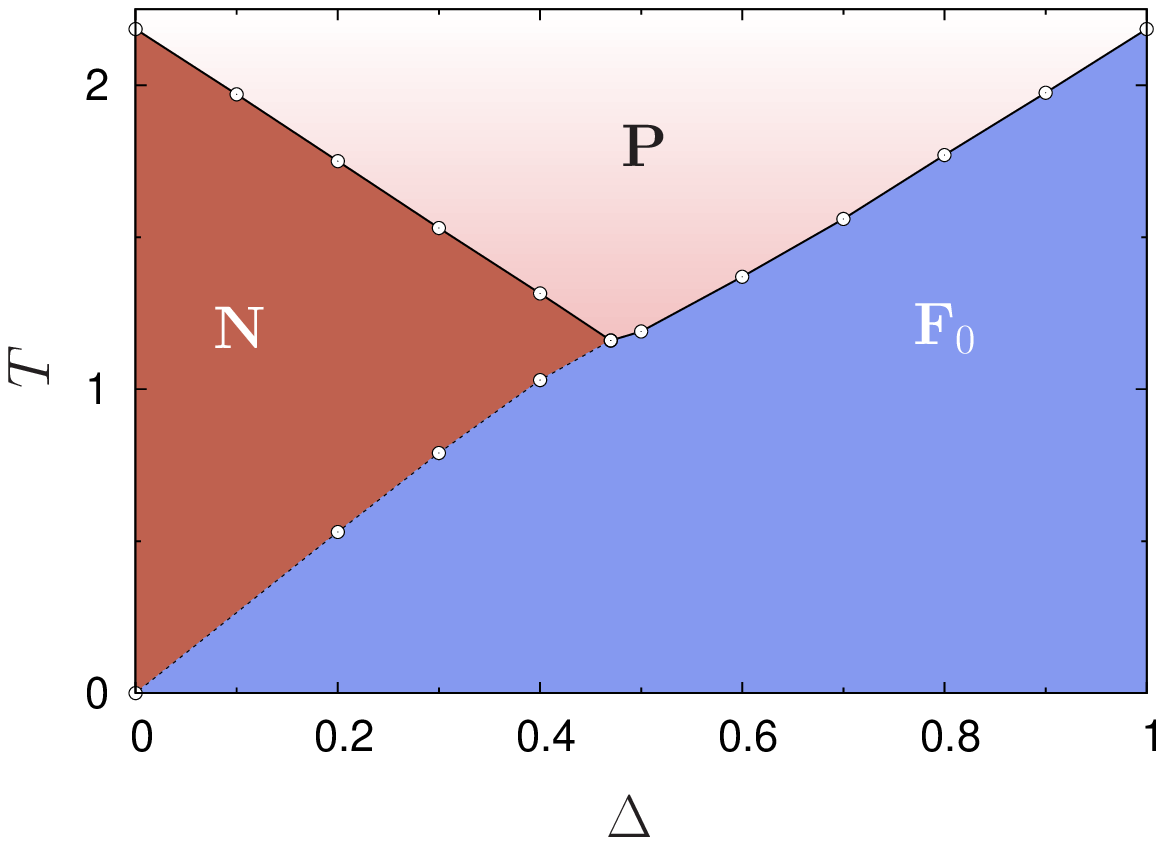}
\includegraphics[width=8cm]{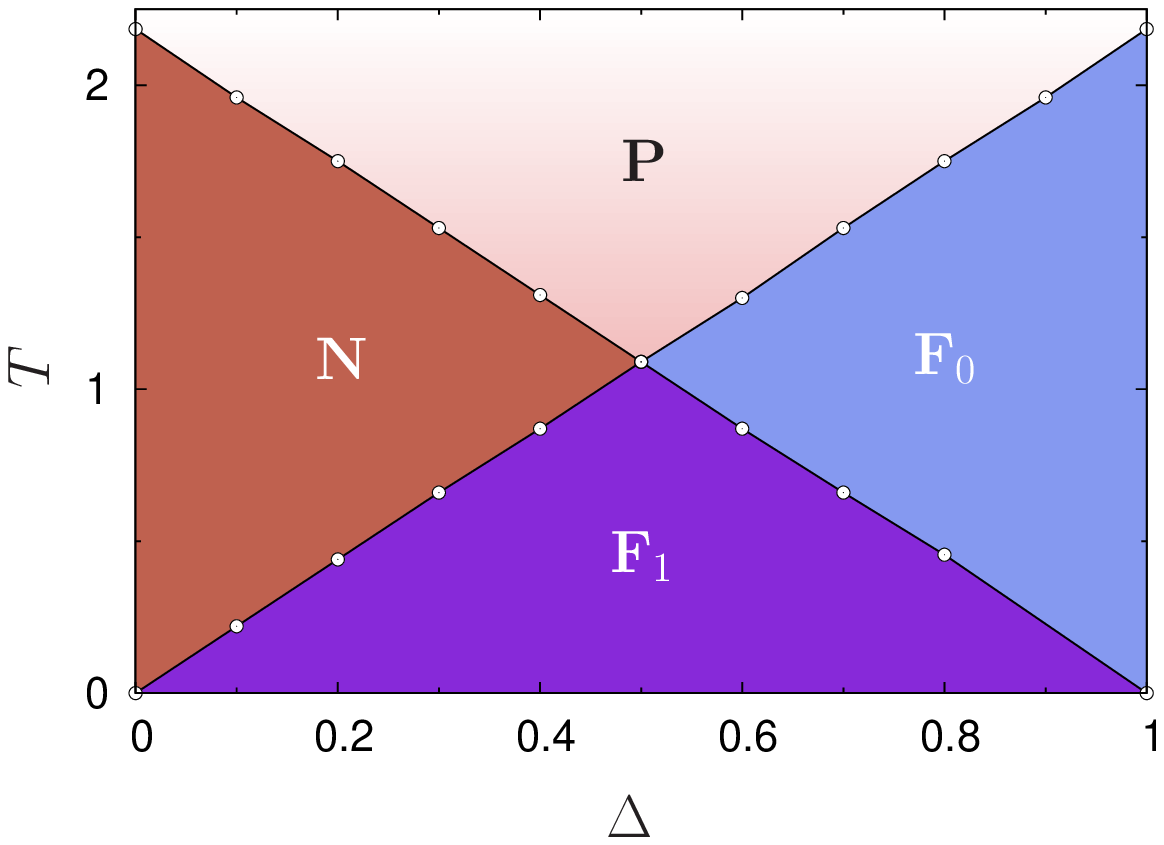}
%
\caption{Phase diagrams of the generalized 3dXY model for $q=3$ (left) 
and $q=8$ (right). The lines are just guide to the eyes and points correspond to the maximum of the specific heat for $L=20$. In both cases, the multicritical point is close to 
$\Delta_{\scriptscriptstyle\rm mc}\simeq 0.5$. At both $\Delta=0$ and 1, the critical temperature is $\tc \simeq 2.202$. With the exception of the {\bf N}-{\bf F}$_0$
transition for $q=3$ that is discontinuous (dashed line), all other transitions are continuous.
As discussed in the text, some uncertainty remains for the $\Delta>0.8$ part of the 
{\bf F}$_1$-{\bf F}$_0$ line.}
\label{diag3Dq3q8}
\end{figure*}

We now address the interesting question of whether the existing ordered 
phases, their splitting as $q$ increases and the nature
of the related transitions are specific to the 2d version of the
model or also occur in other dimensions as well. In
this section, the results obtained for $q=3$ and 8 in 3d are
presented and compared with the 2d case.
Similarly to the latter, the phase diagrams are sketched using the 
specific heat (an example, described later, is shown in Fig.~\ref{fig.3Dq8CvD06} 
for $q=8$ and two values of $\Delta$, 0.6 and 0.8). 
Fig.~\ref{diag3Dq3q8} shows these qualitative phase diagrams for the generalized
3dXY model with $q=3$ (left) and 8 (right).
Instead of being of BKT nature, a common feature of both cases is that
the transitions between the ordered and the paramagnetic phases 
({\bf N}-{\bf P} and {\bf F}$_0$-{\bf P}) are 
second order and belong to the 3dXY universality class, whose critical 
exponents are $\beta\simeq 0.349$, $\gamma\simeq 1.318$, 
$\nu\simeq 0.672$ and $\alpha\simeq -0.015$~\cite{LiTe89,CaHaPeVi06}. 
The exponent $\alpha$ of the specific heat, being negative, indicates the
presence of a cusp instead of a divergence (lambda transition).

\begin{figure}[ht]
\includegraphics[width=8cm]{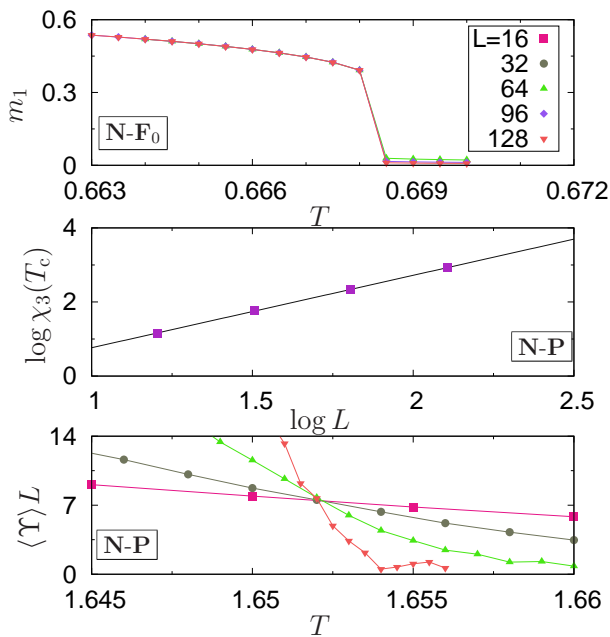}
\caption{Results for the $q=3$ 3dXY model with $\Delta=0.25$. Discontinuous
magnetization {\it vs} temperature around the transition {\bf F}$_0$-{\bf N} (top panel).
At the transition {\bf N}-{\bf P}, the peak of the susceptibility $\chi_3$  
grows as $L^{1.955}$ (middle panel) while all curves for the helicity modulus 
intersect close to the critical temperature (bottom panel).}
\label{3DX3D025_Hel}
\end{figure}

For $q=3$, the phase diagram (Fig.~\ref{diag3Dq3q8}, left) presents the same 
structure in both $d=2$ and 3~\cite{PoArLe11,CaLeAr14}.
The extreme points $\Delta=0$ and 1 are equivalent to 
the original 3dXY model, with the same transition temperature. Moreover, the ground 
state is ferromagnetic for all values of $\Delta$ except $\Delta=0$ where the 
alignment has $\pi/3$ long-range nematic order (although it may be different when
antiferromagnetic interactions are considered~\cite{Zukovic16}). 
In analogy to the 3d three states Potts model, the transition {\bf F}$_0$-{\bf N} is 
discontinuous, as shown by the abrupt jump on 
the magnetization $m_1$ in the top panel of Fig.~\ref{3DX3D025_Hel}. 
On the other hand, the transitions to the paramagnetic phase, from both {\bf N} and
{\bf F}$_0$ phases, are continuous and belong to the 3dXY universality class. 
We show in the middle panel of Fig.~\ref{3DX3D025_Hel}, for $\Delta=0.25$, 
that the susceptibility peak corresponding to $m_3$ at the transition {\bf N}-{\bf P} 
grows as 
$\chi_3(\tc)\sim L^{1.955}$, with $\tc\simeq 1.652$, very close to the expected
$\gamma/\nu\sim 1.96$ value. 
We also measured the helicity modulus whose behavior in the critical region 
is continuous~\cite{FiBaJa73,LiTe89,GoHa93,JeShCh12}
and given by $\Upsilon\sim t^{\upsilon}$, where $\upsilon$ is the critical exponent
and $t$ is the reduced temperature.  
Assuming the universal scaling function $\Upsilon(T,L)=L^{-\upsilon/\nu} g(tL^{1/\nu})$, 
and using the scaling law $\upsilon/\nu=d-2=1$~\cite{FiBaJa73,LiTe89}, $L\Upsilon(\tc,L)=g(0)$ must be independent of the system size at the transition. 
Indeed, the bottom panel of Fig.~\ref{3DX3D025_Hel} shows the rescaled helicity at the transition 
{\bf N}-{\bf P} for $\Delta=0.25$, with all curves intersecting at $\tc\simeq 1.652$, consistent with 
the value obtained via susceptibility.

Fig.~\ref{diag3Dq3q8} (right) shows the phase diagram for $q=8$.  
Instead of the three ferromagnetic-like phases in the 2d case, only two remain 
in 3d (we denote the intermediate phase by {\bf F}$_1$).  
As mentioned before, the transition to the paramagnetic phase, from both {\bf N} and
{\bf F}$_0$, are in the 3dXY universality class.
 We now discuss the properties of the transitions involving
the intermediate phase, {\bf F}$_1$-{\bf N} and {\bf F}$_1$-{\bf F}$_0$, and
the evidences for a crossover inside the phase {\bf F}$_1$.

\begin{figure}[ht]
\includegraphics[width=8cm]{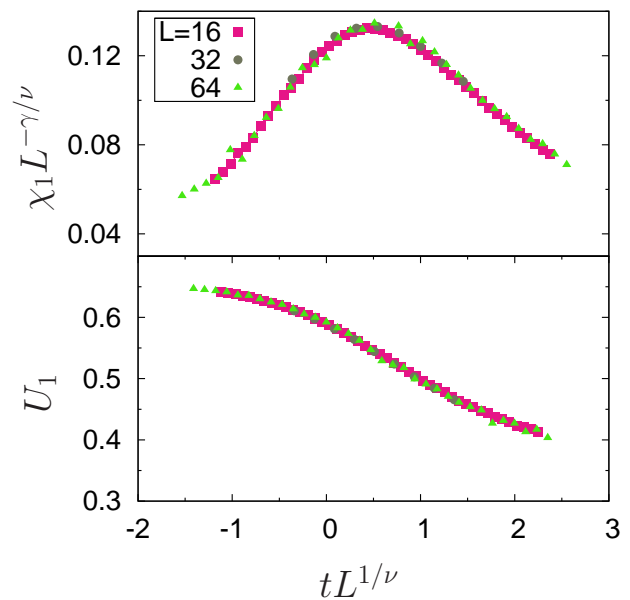}
\caption{Collapses of the susceptibility (top) and Binder cumulant (bottom) at the 
{\bf F}$_1$-{\bf N} transition for $q=8$ and $\Delta=0.35$ with $\tc=0.771$ and the 
exponents $\gamma\simeq 1.33$ and $\nu\simeq 0.67$, whose values are close to those of the 3dXY model.}
\label{3DX1D035}
\end{figure}

Fig.~\ref{3DX1D035} (top) shows the behavior of the susceptibility $\chi_1$ for $\Delta=0.35$
around the {\bf F}$_1$-{\bf N} transition. An excelent data collapse is obtained with
the critical exponents $\gamma\simeq 1.327$, $\nu\simeq 0.671$ and $\tc\simeq 0.771$, 
values that are very close to those of the 3dXY universality class. 
Moreover, the Binder cumulant, evaluated for different system sizes, has the typical crossing
point and the rescaled data also collapses very well onto a universal curve 
using the same $\tc$ and $\nu$ above, as shown in Fig.~\ref{3DX1D035} (bottom).
This is remarkable since in principle one would expect a symmetry breaking, 3d Ising 
universality class transition. Indeed, the angle distribution in these two phases is similar in both two
and three dimensions (see, e.g., Fig.~\ref{fig.histq8}), passing from equally
distributed peaks around the circle (nematic-like) to a few peaks on a single
half-plane (ferromagnetic-like). Moreover, the helicity does not present the 
typical signature of the 3dXY transition, continuously transitioning to the value
characterizing the new phase, with an intermediate, size dependent behavior. The specific
heat is of no help to decide between those universality classes, a good collapse
is obtained with the above $\nu$ (closer to 3dXY) and a small but positive $\alpha$
(closer to 3d Ising). Therefore, our present data only partially confirm
that the transition is in the 3d Ising universality class, while larger sizes 
and corrections to the scaling will be necessary to obtain a better estimate of
the exponents.

\begin{figure}[ht]
\includegraphics[width=8cm]{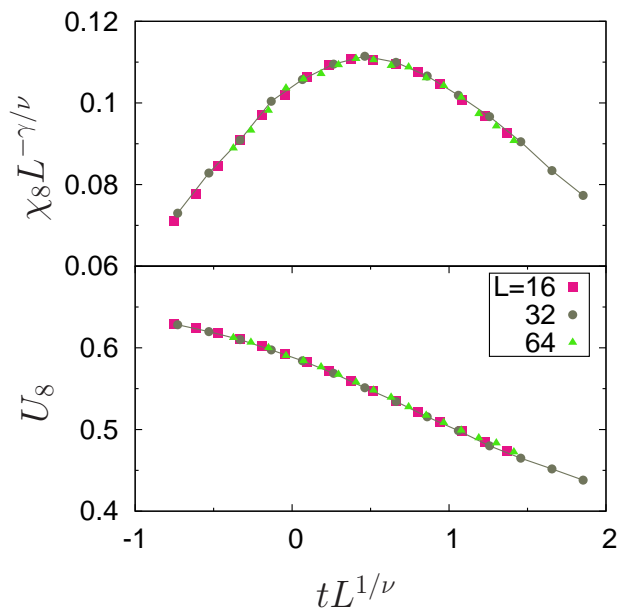} 
\caption{{\bf F}$_1$-{\bf F}$_0$ transition for $q=8$ and $\Delta=0.6$ in 3d.
(Top) Susceptibility {\it vs} temperature for the 3d 
collapses onto a universal curve with $\tc=0.881$ and exponents 
$\gamma\simeq 1.33$, $\nu\simeq 0.67$. (Bottom) Collapse of the Binder cumulant using the same values.}
\label{3DX8D06}
\end{figure}

%

\begin{figure}[ht]
\includegraphics[width=8cm]{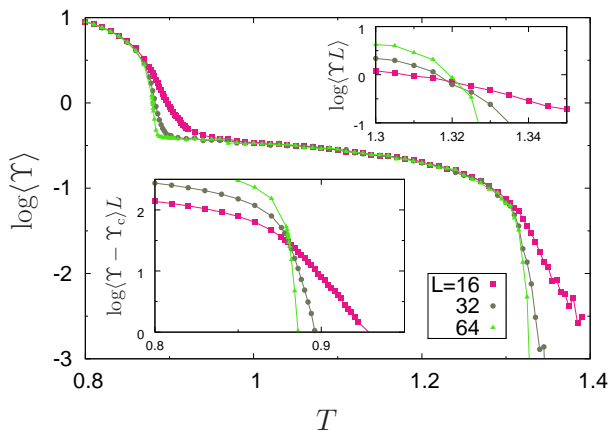}
\caption{Helicity modulus versus temperature in 3d for $q=8$, $\Delta=0.6$ and
several system sizes. The two transitions can be seen as two jumps, first
to a finite value ($\Upsilon_{\scriptstyle\rm c}\simeq 0.404$) at the {\bf F}$_1$-{\bf F}$_0$ transition
($\tc\simeq 0.88$) and then to zero at the {\bf F}$_0$-{\bf P} transition 
($\tc\simeq 1.32$). At both transitions, the helicity, conveniently rescaled,
crosses at the critical temperature for different values of $L$, as shown in the
two insets, bottom and top, respectively.}
\label{hel3Dq8D06}
\end{figure}

\begin{figure}[ht]
\includegraphics[width=8cm]{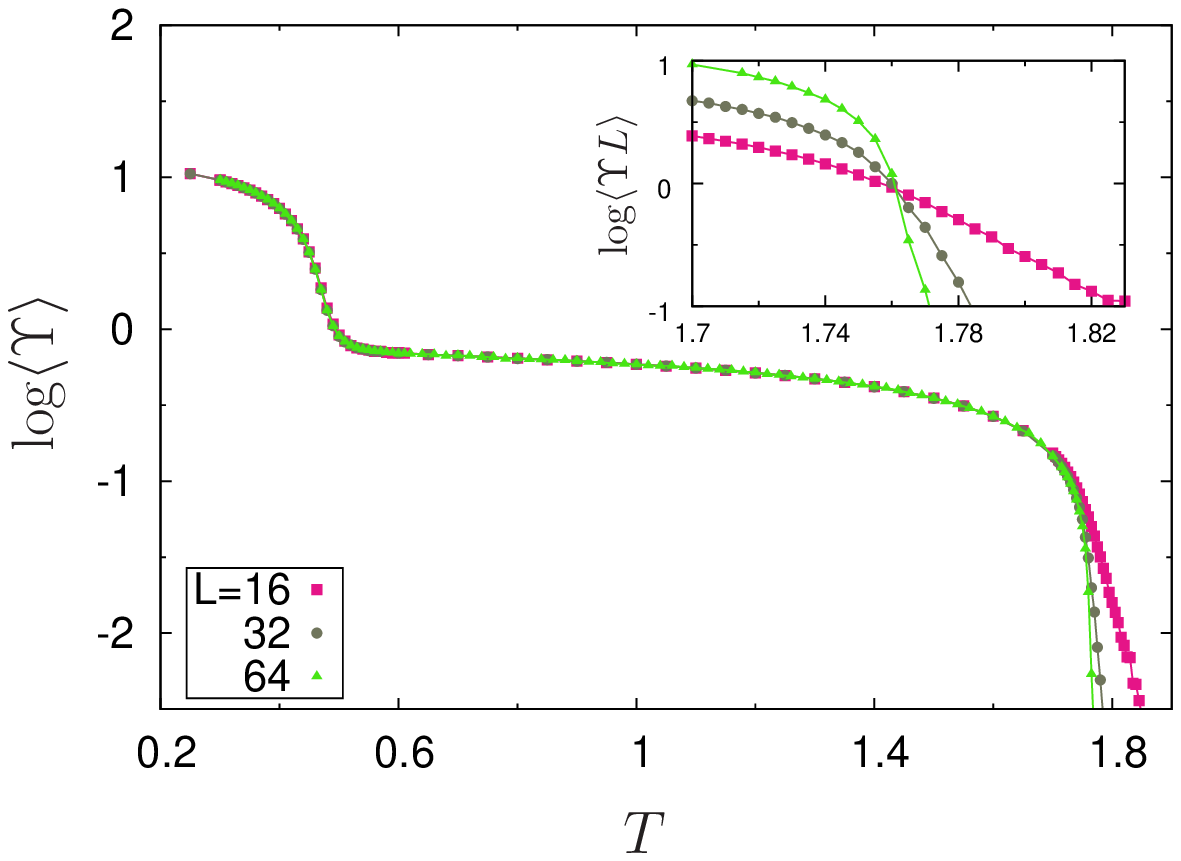}
\caption{The same as in Fig.~\ref{hel3Dq8D06} but with $\Delta=0.8$. Notice that although
the helicity has a sudden decrease around $T\simeq 0.4$, it does not present any
size dependence.}
\label{hel3Dq8D08}
\end{figure}

The transition {\bf F}$_1$-{\bf F}$_0$ also seems to be in the 3dXY universality class (but
see below).
Indeed, Fig.~\ref{3DX8D06} (top) shows the collapsed susceptibility $\chi_8$ around 
$\tc\simeq 0.88$, for $\Delta=0.6$,  
with the critical exponents $\gamma\simeq 1.33$ and $\nu\simeq 0.67$, once again
close to those of the 3d XY univesality class. At this critical temperature, the
Binder cumulant also presents a crossing point for several system sizes and in
the bottom part of Fig.~\ref{3DX8D06} its collapse using the same exponent 
$\nu$ and $\tc$ is shown. 
Fig.~\ref{hel3Dq8D06} shows the helicity modulus for the same $\Delta$. Inside each phase
it has a size independent value, with all curves collapsing onto each other, as in 2d. 
At the two
transitions ({\bf F}$_1$-{\bf F}$_0$ and {\bf F}$_0$-{\bf P} at $\tc\simeq 0.88$
and 1.32, respectively) there is a continuous change of $\Upsilon$ towards the
value at the transition ($\Upsilon_{\scriptstyle\rm c}\simeq 0.404$ and 0,
respectively). From the previous discussion of $\Upsilon$, we expect a power
law behavior at the transition, $\Upsilon - \Upsilon_{\scriptstyle\rm c}
\sim t^{\upsilon}$, with $\upsilon=\nu$. A good collapse (not shown)
is obtained with values very close to the $\tc$ and $\nu=\upsilon$ obtained 
from the susceptibility. Moreover, the specific heat for this case is
shown in the top pannel of Fig.~\ref{fig.3Dq8CvD06}. The first peak on the
left is believed to correspond to a crossover (see below) while the rightmost
one is associated to the transition to the paramagnetic state. The behavior of 
the intermediate specific heat peak, instead, is consistent with a lambda 
transition and, indeed, $\alpha<0$ for the 3d XY universality class.

The results are less clear at the far right region of the phase diagram and how
and whether the {\bf F}$_1$-{\bf F}$_0$ transition line extends beyond this
point is still an open problem. For
example, for $\Delta=0.8$ (similar results were obtained for
$\Delta=0.9$ as
well), the specific heat in Fig.~\ref{fig.3Dq8CvD06} (bottom)
has only two peaks, one that clearly corresponds to the lambda transition at higher temperatures
while the other one, at lower temperature, does not present any size dependence on the 
range of system sizes considered
here. The same behavior is observed in the helicity, Fig.~\ref{hel3Dq8D08}: although
two sudden decreases are observed as the temperature increases, only the one
at the highest $T$ presents a size dependence compatible with a continuous transition
(inset). The other one, corresponding to the first peak of the specific heat, does not 
change with the system size. Notice that, as discussed below, there is a crossover
line inside phase {\bf F}$_1$ that meets the {\bf F}$_1$-{\bf F}$_0$ close to
$\Delta=0.8$, what may perhaps explain the odd behavior in this case. 

\begin{figure}[ht]
\includegraphics[width=8cm]{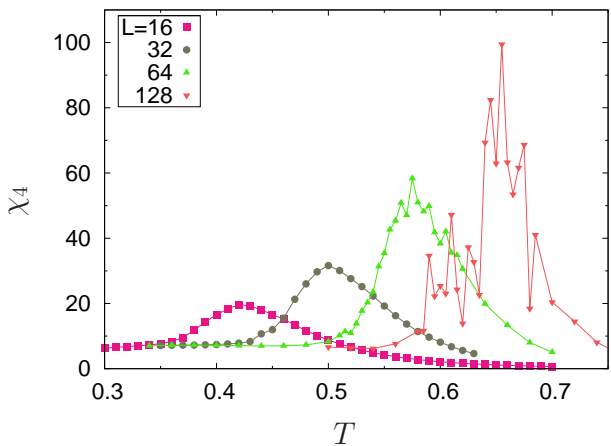}
\caption{Susceptibility $\chi_4$ {\it vs} temperature for $\Delta=0.4$ and $q=8$
in 3d showing that, for small system sizes, it develops a peak at a temperature 
below the {\bf F}$_1$-{\bf F}$_0$ transition (that, in 2d, corresponds to the actual
transition {\bf F}$_1$-{\bf F}$_2$). However, as $L$ increases, this peak moves to 
the right, towards the {\bf F}$_2$-{\bf N} boundary at $\tc\simeq 0.88$.}
\label{3DX4D04}
\end{figure}

Finally, we investigate the possible existence of a phase {\bf F}$_2$ similar to the
2d case. Small systems behave as if another transition indeed exists inside the 
{\bf F}$_1$ phase: both the specific heat (Fig.~\ref{fig.3Dq8CvD06} for 
$\Delta=0.6$) and the susceptibility $\chi_4$ (Fig.~\ref{3DX4D04} for $\Delta=0.4$) 
present a peak in that region.
However, by increasing the system size, all the evidence points to a crossover instead 
of a genuine transition. While the specific heat peak of Fig.~\ref{fig.3Dq8CvD06}
(top) does not change in height, the susceptibility in Fig.~\ref{3DX4D04} does
increase in size while, at the same time, moving to higher temperatures. With our
present data, it seems that this line moves toward the boundary with the {\bf N} phase,
a pure crossover.

\section{Conclusions}
\label{conclusion}

We performed extensive simulations for a generalization of the
continuous spin XY model that introduces competition between different 
local alignments. Despite both terms in the Hamiltonian, Eq.~(\ref{eq.H}), 
having the same symmetry under uniform rotations, each one induces 
a different local ordering: while the ferromagnetic term tends to align
the spins, the nematic one tries to have them either parallel or
dephased by the multiples of $2\pi/q$ (a nematic configuration). 
While the former configuration is favored by both terms, the latter has 
some degree of frustration and only appears at  
higher temperatures, whatever the value of $q>1$. The
low temperature phase, on the other hand, has a local ferromagnetic
ordering that, in both 2 and 3 dimensions, unfolds through a sequence 
of phase splittings, as $q$ increases, into several phases with
similar ferromagnetic ordering. As a consequence, for $q\geq 4$ this 
model has intermediate phases ($0<\Delta<1$) driven by the competition 
of both terms in Eq.~(\ref{eq.H}), except for $\Delta=0$ and 1 where 
a single phase exists, with either local nematic or ferromagnetic alignment,
respectively. It is interesting that the appearance of a new phase
below the nematic one, and the subsequent splitting and growth is
very similar to the sequence appearing in the discrete, $\mathbb{Z}_q$ 
symmetry Clock model. 
Indeed, in 2d the transition to phase {\bf N} is in the Ising ($q=2$ and 4) 
or 3 states Potts model ($q=3$) universality class. All phases below the paramagnetic 
transition are critical in the BKT sense (e.g.,
power-law decaying correlations and divergent susceptibility, with non
universal exponents, at all temperatures). Associated with that, the model
presents multiple transitions as the temperature changes, some
with the BKT signature, while others are discrete symmetry breaking transitions 
embedded into these critical regions. 
These phases may contain integer and/or fractional vortices, which can be bound
or unbound. 
While for $q=2$ and 3 the low temperature phase
is dominated by the integer vortices of phase {\bf F}$_0$, for $q=8$ (and
probably for any $q>3$), these integer vortices coexist, in the competition 
induced phases, with fractional vortices and topological defects at sufficiently
low temperatures. 
Remarkably, having vortices decoupling at a transition is not a
suficient condition for this transition to be of BKT type. Indeed, as an
example, entering the {\bf F}$_0$ phase (whose bound vortices are
integer) from {\bf F}$_1$ or
{\bf F}$_2$ (both populated with two species of bound vortices) may
be either a BKT or an Ising transition, respectively. Moreover, for a second order
transition, decoupling is not even necessary as it does not occur at 
the {\bf F}$_1$-{\bf F}$_2$ transition.  
We may also view the interplay between the two terms in 
Eq.~(\ref{eq.H}) as a way to choose which species of vortices to 
suppress or enhance~\cite{JeWe92,BhRa16} 
by tuning $\Delta$ and $T$.

A BKT transition is signaled by a discontinuity in the helicity modulus that,
for a finite system, appears as a size dependence on the intersection
of $\Upsilon$ with a given reference line. When there was a sudden, significant
change in the helicity, but finite size effects were
absent, we assumed it to be an indication that the transition is
second order (another possibility, one that occurs in the 3d case, is a crossover). 
However, the true nature of some of the transitions reported 
here is only observed for systems too large to be simulated on a single CPU. 
It was only through the power made available by GPU processing that we were 
able to obtain our results, and even so, this characterization is not always very 
clear, with a few regions, both in two and three dimensions, that deserve further
studies. In the case of the {\bf F}$_1$-{\bf F}$_2$ transition occurying for
$q=8$ in 2d, not even this extra computational aid was sufficient. 
Further work is therefore necessary in order to make sure that the transition
is real and if so, obtain more precise estimates for the critical exponents, 
deciding whether they depend or not on $\Delta$, as the results so far indicate.
In the 3d case, although there seems to exist a 
single intermediate phase as the 2d {\bf F}$_1$-{\bf F}$_2$ transition turns
into a crossover, there is some uncertainty regarding the extension of the
{\bf F}$_1$-{\bf F}$_0$ line beyond $\Delta\simeq 0.8$ as several quantities
behave qualitatively different from the $\Delta=0.6$ case (compare, e.g., the
first shoulder in Figs.~\ref{hel3Dq8D06} and \ref{hel3Dq8D08}). It would take
larger systems and longer simulation times in order to properly access these
issues, what is beyond our current computational capabilities.

Recently, a new universality class has been studied~\cite{HuBaBe15} in models where 
the BKT transition meets a $\mathbb{Z}_2$ (Ising) transition line at a multicritical 
point. This point seems to have properties of supersymmetry. 
We conjecture that this may also be the case for the multicritical point exibited by the
generalized XY model in 2d with $q=2$, that presents both $U(1)$ and 
$\mathbb{Z}_2$ symmetry properties. It is thus of interest to
study not only the properties of the multicritical point in this model, but
also to see whether it conforms with the predictions of Ref.~\cite{HuBaBe15}
and whether this applies as well for $q=3$ where the line is in the Potts universality
class or for larger values of $q$. Moreover, for $q>4$ there is a second multicritical point that is at the
triple border between the three {\bf F}$_i$ phases. The behavior at this point and
whether it is similar to the other one is an yet open problem. It would be
interesting to study the properties of these multicrical points in order
to check whether the behavior of the correlation length differs both from a conventional
second order transition and from a BKT one as predicted in Ref.~\cite{HuBaBe15}. 
Further possible extensions include 
non integer values of $q$, antiferromagnetic interactions~\cite{ZuId03} or a generalized 
version with discrete spins (studied for $q=2$ in Ref.~\cite{Chatelain16}). Trying to
disentangle the roles of the two terms in the Hamiltonian, by introducing two variables
per site and a on-site coupling between them\cite{BrAe82,GrKo86,GhSh05,ShJi05,PaRaRo16}, may also be helpful to better understand
the new phases for $q>4$. These questions, together with
a better description of the geometry of topological defects 
are left for a future work.

\begin{acknowledgments}

Research partially supported by the Brazilian agencies CNPq, CAPES and Fapergs.
JJA acknowledges the INCT-Sistemas Complexos (CNPq) for partial support and
interesting discussions with M. Picco and F. Corberi.
\end{acknowledgments}

\bibliographystyle{apsrev4-1}
%

\end{document}